\documentclass[traditabstract,longauth]{aa} 
\usepackage{graphicx}
\usepackage{txfonts}
\usepackage{natbib}

\def\degr{\hbox{$^\circ$}}

\def\arcsec{\hbox{$^{\prime\prime}$}}

\begin{document}

   \title{The Pipe Nebula as seen with \emph{Herschel}:\\
   Formation of filamentary structures by large-scale compression\thanks{\emph{Herschel} is an ESA space observatory with science instruments provided by European-led Principal Investigator consortia and with important participation from NASA.} ?}


   \author{N. Peretto,
          \inst{1}
          Ph. Andr\'e,
          \inst{1}
         V. K\"onyves\inst{1}, 
         N. Schneider\inst{1}, D. Arzoumanian\inst{1}, P. Palmeirim\inst{1}, P. Didelon\inst{1},  M. Attard\inst{1}, J.P. Bernard\inst{2}, J. Di Francesco\inst{3}, D. Elia\inst{4}, M. Hennemann\inst{1}, T. Hill\inst{1}, J. Kirk\inst{5}, A. Men'shchikov\inst{1}, F. Motte\inst{1}, Q. Nguyen Luong\inst{1}, H. Roussel\inst{6}, T. Sousbie\inst{6}, L. Testi\inst{7}, D. Ward-Thompson\inst{8}, G. J. White\inst{9,10} A. Zavagno\inst{11}
          }

   \institute{\inst{1} Laboratoire AIM, CEA/DSM-CNRS-Universit\'e Paris Diderot, IRFU service d'Astrophysique, C.E. Saclay, Orme des merisiers, 91191 Gif-sur-Yvette, France \\
    \inst{2} Universit\'e de Toulouse, UPS, CESR, 9 avenue du colonel Roche, 31028 Toulouse Cedex 4, France \\
    \inst{3} National Research Council of Canada, Herzberg Institute of Astrophysics, University of Victoria, Department of Physics and Astronomy, Victoria, Canada\\
    \inst{4} INAF - Instituto Fisica Spazio Interplanetario, via Fosso del Cavaliere 100, 00133 Roma, Italy\\
    	\inst{5} School of Physics and Astronomy, Cardiff University, Queens Buildings, The Parade, Cardiff, CF243 AA, UK\\
   \inst{6} Institut d'Astrophysique de Paris and Universit\'e Pierre et Marie Curie (UPMC), UMR 7095 CNRS, 98 boulevard Arago, 75014, Paris, France\\
   \inst{7} ESO, Karl Schwarzschild str. 2, 85748 Garching bei Munchen, Germany\\
   \inst{8} University of Central Lancashire, Jeremiah Horrocks Institute, PR12HE, UK\\
   \inst{9} Department of Physical Sciences, The Open University. Milton Keynes MK7 6AA, England\\
   \inst{10} RALSpace, The Rutherford Appleton Laboratory, Chilton, Didcot, OX11 0NL, England\\
   \inst{11} Laboratoire d'Astrophysique de Marseille UMR6110, CNRS, Universit\'e de Provence, 38 rue F. Joliot-Curie, 13388 Marseille, France\\
              \email{Nicolas.Peretto@cea.fr}
             }
             
   \date{Received ; accepted }

 
\abstract
   {A growing body of evidence indicates that the formation of filaments in interstellar clouds is a key component of the star formation  process. In this paper, we present new \emph{Herschel} PACS and SPIRE observations of the B59 and Stem regions in the Pipe Nebula complex, revealing a rich, organized network of filaments. The asymmetric column density profiles observed for several filaments, along with the bow-like edge of B59, indicates that the Pipe Nebula is being compressed  from its western side, most likely by the winds from the nearby Sco OB2 association. We suggest that this compressive flow has contributed to the formation of some of the observed filamentary structures.  In B59, the only region of the entire Pipe complex showing star formation activity, the same compressive flow has likely enhanced the initial column density of the clump, allowing it to become globally gravitationally unstable. Although more speculative, we propose that gravity has also been responsible for shaping the converging filamentary pattern observed in B59. While the question of the relative impact of large-scale compression and gravity remains open in B59,  large-scale compression appears to be a plausible mechanism for the initial formation of filamentary structures in the rest of the complex.}
   
\authorrunning{Peretto et al.}
\maketitle
%

\section{Introduction}

While stars form from prestellar cores, cores themselves form from larger structures, often having filamentary shapes. Understanding the physical connection between cores and associated filaments is central to establishing a broader picture of star formation in the Galaxy. Recent observations with the \emph{Herschel} space observatory \citep{pilbratt2010} have revealed the ubiquity of complex networks of filaments in nearby interstellar clouds \citep{andre2010}, as well as in more distant and massive clouds \citep{molinari2010,hill2011}. The unprecedented sensitivity of both \emph{Herschel} imaging cameras PACS \citep{poglitsch2010} and SPIRE \citep{griffin2010} facilitates the systematic detection of filaments down to A$_{\rm V}\simeq 0.1$ in amplitude.
\cite{arzoumanian2011} characterized the physical properties (widths and column densities) of such filaments in  three Gould Belt regions,
and found a relatively uniform filament width of $\sim0.1$~pc (FWHM), independent of their lengths, or column densities. 
The origin of such a uniform filament width remains unclear. \citet{arzoumanian2011} proposed that it could be a consequence of the process of filament formation by the large-scale turbulent compression of interstellar material. Only those filaments reaching the critical mass per unit length M$_{line,crit} = 2 c_s^2/G$ \citep{ostriker1964} would fragment into prestellar cores, and finally form stars \citep{andre2010}.  In standard molecular clouds, the isothermal sound speed c$_s \simeq 0.2$~km\,s$^{-1}$ (for T$_k=10$~K) leads to  M$_{line,crit} \sim 15$~M$_{\odot}$\,pc$^{-1}$ (or A$_{\rm V} \simeq 8$ for filaments of $\sim0.1$~pc width).

Here,  we present new \emph{Herschel} observations of the Pipe Nebula, one of the closest known star-forming regions. With a distance of $d=145$~pc \citep{alves2007} and only a few identified young stellar objects \citep[YSOs - e.g.][]{forbrich2009}, the Pipe Nebula is a pristine example of a nearby cloud, and an ideal target for the study of filament and core formation. It is mainly composed of an elongated ($L\simeq18$~pc) dark cloud subdivided into three parts: B59 and the Stem, which are discussed here,  and the Bowl, which will be presented in a forthcoming paper. B59 is the only region showing clear signs of star formation activity \citep{onishi1999,forbrich2009}, with a small cluster of YSOs located right in its center. 
Measurements of the morphology of  the magnetic field  using optical dust polarisation have also been performed and showed that its direction is mainly orthogonal to the main body of the Pipe \citep{alves2008,franco2010}. Using the indirect method of Chandrasekhar \& Fermi, these authors  estimated a magnetic field strength  of a few tens of  $\mu$G.  While these authors claimed that the Pipe Nebula structure has been mostly shaped by the magnetic field, \citet{heitsch2009} proposed that gravity alone drives the evolution of the entire complex. 
These two examples illustrate the controversy about  the star formation processes operating in this interesting region. 


 \begin{figure*}
   \vspace{-0cm}
   \hspace{0cm}
   \includegraphics[width=18cm,angle=0]{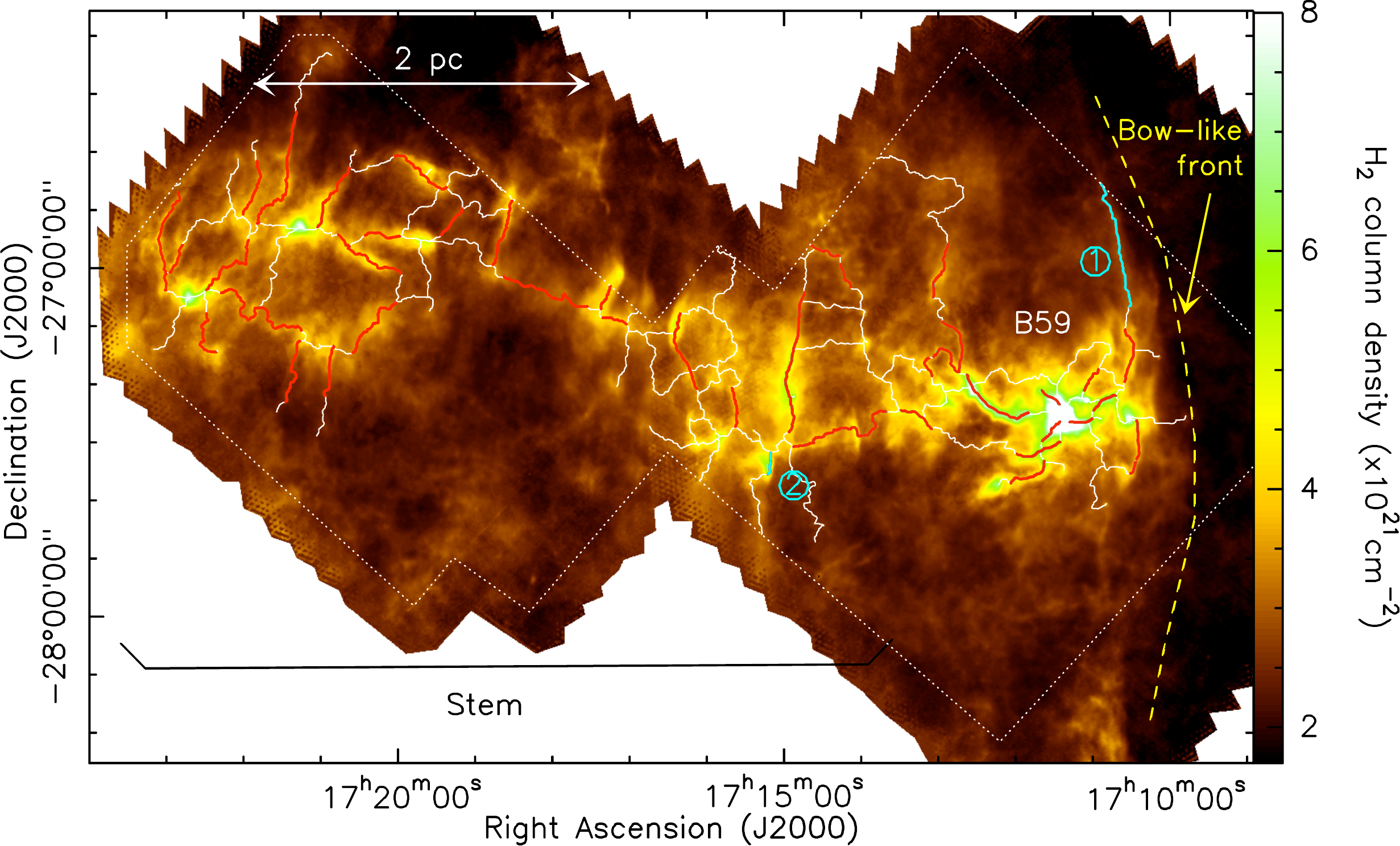}
   \vspace{-0cm}
      \caption{\emph{Herschel} column density map of the Pipe on which is overlaid the filament network as traced with the DisPerSE algorithm \citep{sousbie2011}. The 43 filaments selected after eye inspection are displayed in red and cyan while the remaining filaments are displayed in white.  Those presented in cyan and indicated with numbers are shown in more detail in Fig.~\ref{filprof}. The white dotted polygon marks the area covered by both PACS and SPIRE imaging cameras. The yellow dashed line emphasizes the bow-like front of the B59 clump. The angular resolution of this image is 36\arcsec\ (HPBW). In this paper, we define the Stem the area of our \emph{Herschel} map located above R.A.$=17^h13^m30^s$, while B59 corresponds to the region displayed in online Fig.~\ref{zoom}.
              }
         \label{skel}
   \end{figure*}

\section{\emph{Herschel} imaging observations}

We present SPIRE/PACS parallel mode observations of the Pipe Nebula obtained 
as part of  the {\it Herschel} Gould Belt Survey (GBS; Andr\'e et al. 2010). This observing mode allows the simultaneous observations in 5 bands centered at 70/160 $\mu$m for PACS and 250/350/500 $\mu$m for SPIRE. The {\it Herschel} GBS Pipe data consist of three fields\footnote{For more details cf. http://gouldbelt-herschel.cea.fr} roughly corresponding to the three parts of this dark cloud (B59, Stem, and Bowl, see Sect.~1);  only those corresponding to the Stem and B59 regions are presented here.
The PACS data  were reduced with improved calibration within HIPE\footnote{HIPE is a joint development software by the \emph{Herschel} Science Ground Segment Consortium, consisting of ESA, the NASA \emph{Herschel} Science Center, and the HIFI, PACS and SPIRE consortia} 7.0. After standard HIPE data reduction steps, up to level 1, the maps were produced with v10 of Scanamorphos\footnote{{\it http://www2.iap.fr/users/roussel/herschel/}} \citep{roussel2012}. The SPIRE data were reduced with HIPE 7.1956, using modified pipeline scripts. Data taken during the turnarounds at the map borders were included, and a destriper module with a polynomial baseline of order zero was applied. The two cross-linked coverages were then combined using the default 'naive-mapper' to produce the final map.


\section{Column density distribution}

\begin{figure*}
   \vspace{-0cm}
   \hspace{-0cm}
   \includegraphics[width=7.8cm,angle=0]{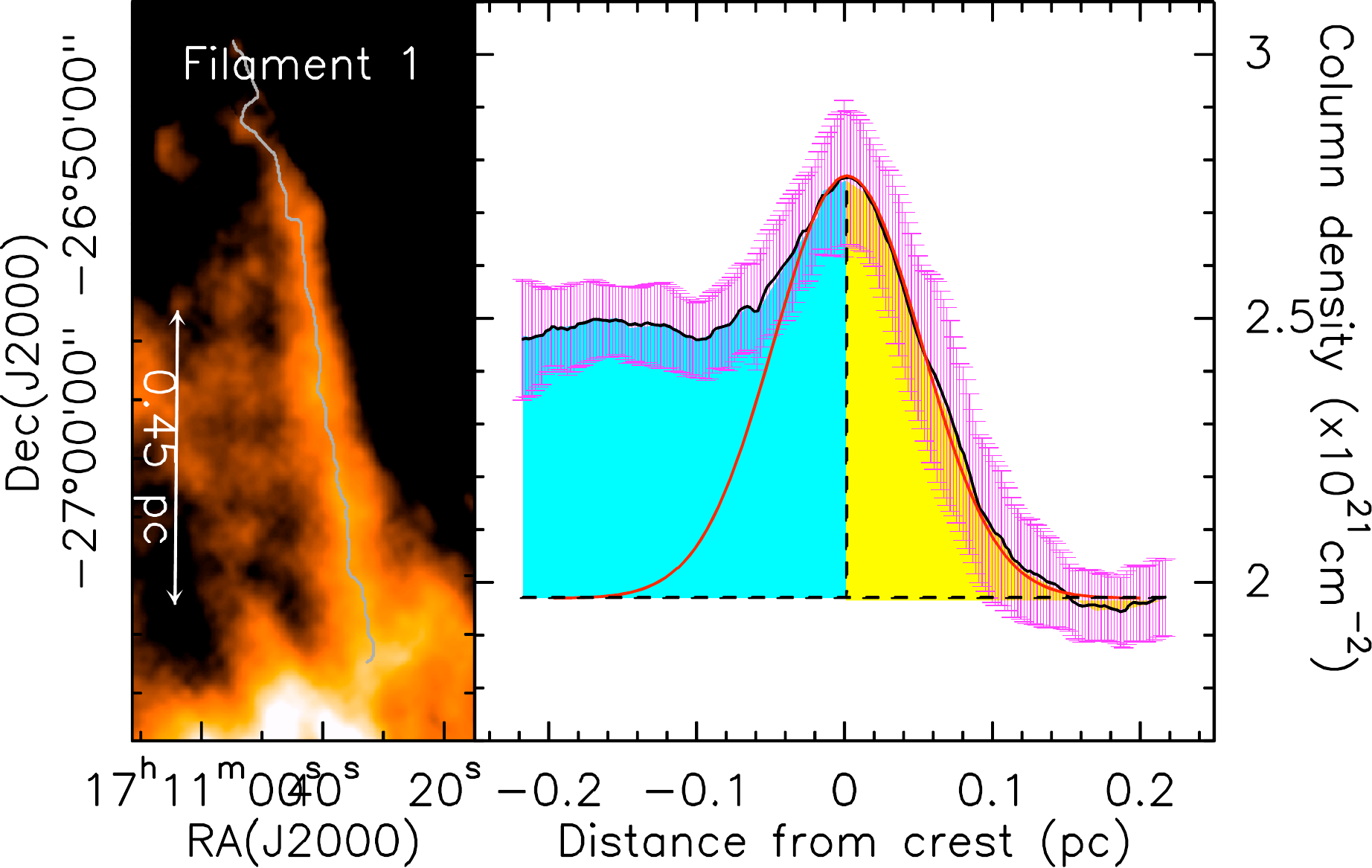}
   \hspace{0.2cm}
   \includegraphics[width=10cm,angle=0]{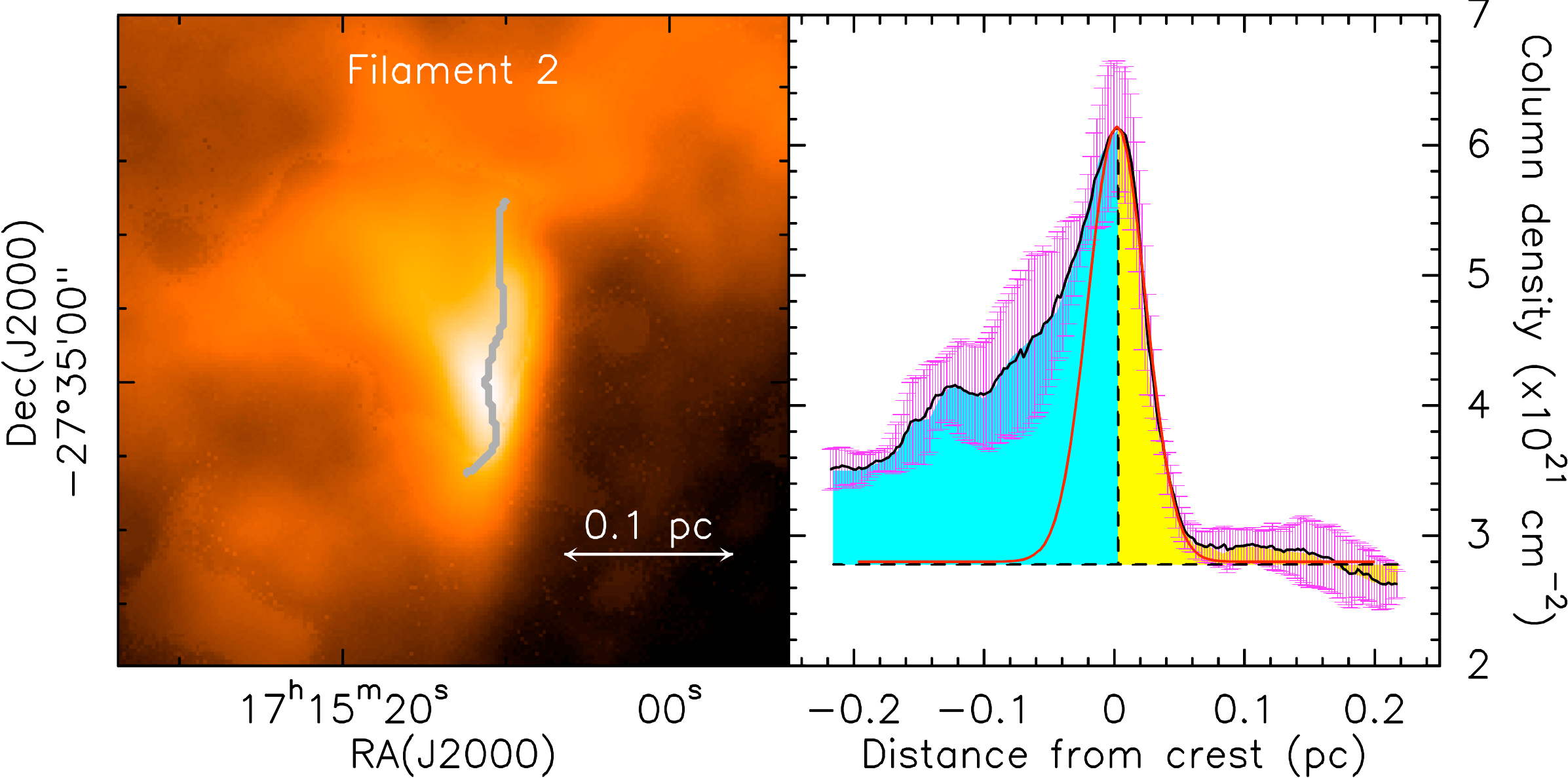}
   \vspace{-0cm}
      \caption{Two examples of filamentary structures with asymmetric transverse profiles. For both filaments, we show on the left an enlargement of the column density map over which is plotted the crest of the filament, and on the right the averaged column density profile measured perpendicular to the crest. Filament 1 corresponds to the bow-like edge of B59, while Filament  2 corresponds to a filament located in the Stem of the Pipe complex (see dark blue filaments in Fig.~\ref{skel}). The blue and yellow shaded areas show the regions that have been integrated to quantify the observed asymmetry. The Gaussian fits performed to infer the FWHM widths of the two filaments are shown as red solid lines.
              }
         \label{filprof}
   \end{figure*}

An important output of  the {\it Herschel} parallel mode imaging observations are column density maps. We recovered the {\it Herschel} zero-flux levels of the Pipe fields by correlating the \emph{Herschel} data with Planck and IRAS data \citep{bernard2010}. 
To construct column density maps, we performed pixel-by-pixel grey-body 
fitting from 160 $\mu$m to 500 $\mu$m, fixing the specific dust opacity such that $\kappa_{\nu}= 0.1\times\left(\frac{\nu}{1000~GHz}\right)^2$~cm$^2$\,g$^{-1}$ \citep[cf.][]{hildebrand1983} and leaving the H$_2$ column density and dust temperature as free parameters. This calculation was done everywhere PACS and SPIRE data were both available. Due to coverage differences between the two instruments in parallel mode observations, a small fraction of the field (on the edge - see Fig.~\ref{skel}) has only SPIRE data. For these pixels, we used a single temperature $T_{dust}=15.4$~K, i.e, the median temperature in the field, along with the 500 $\mu {\rm m}$ data to recover the column density structure. The resulting 36\arcsec\ resolution column density map is shown as a background image in Fig.~\ref{skel}  (see also online Fig.~\ref{cdmap}). On this Figure, the densest region in the western part corresponds to the well-known B59 cluster-forming clump \citep[e.g.][]{covey2010}, while the rest of the map corresponds to the Stem. Both regions show a rich network of filaments and cores. An interesting feature of this map is the sharp, 3 pc long, bow-like edge on the western side of B59 which  is associated with a$\sim 30\%$ increase of the background column density between either side of this edge (see Fig.~\ref{filprof}).
Also, note that only a small fraction (i.e. 0.6\%) of the mapped region corresponds to material above $A_{\rm V}\simeq8$ (see online Fig.~\ref{cdmap}, where we use the conversion factor from \citet{bohlin1978}: N$_{H_2}\simeq 0.94\times A_V \times 10^{21}$~cm$^{-2}$), the extinction threshold above which prestellar core and protostar formation is believed to take place \citep[e.g.][]{heiderman2010,andre2010}.

 

\section{Analysis of the filamentary structure}

\subsection{Filament selection}

To characterize the filamentary structure of the Pipe,  we used the DisPerSE algorithm \citep{sousbie2011} to trace the central crests of all filaments having A$_{\rm V} \ge 2$ (see Fig.~\ref{skel}). 
The goal here is to select a set of filaments that we can directly see by eye in the \emph{Herschel} column density map, the question of the completeness of the filament extraction procedure being beyond the scope of this paper. For this reason, we inspected by eye the filament network obtained with DisPerSE from the Pipe column density map at 36\arcsec\ resolution (i.e. 0.026~pc). We visually selected a final sample of 43 filaments considered as reliable based on their contrast relative to their immediate surrounding background.  Although this small sample is not meant to be statistically complete, we checked that the position angle distribution (measured east of north, see Appendix) of the selected filaments is representative of the full network. Online figure~\ref{histo_pa_all} shows  both distributions. A Kolmogorov-Smirnov test on these indicates that there is a probability of $\sim87\%$ that the corresponding filament samples have the same parent distribution. We are thus confident that the conclusions of the paper do not depend on this filament selection.

\subsection{Organization and transverse profiles of the filaments}

\begin{figure}
   \vspace{-0cm}
   \hspace{-0cm}
   \includegraphics[width=7.8cm,angle=0]{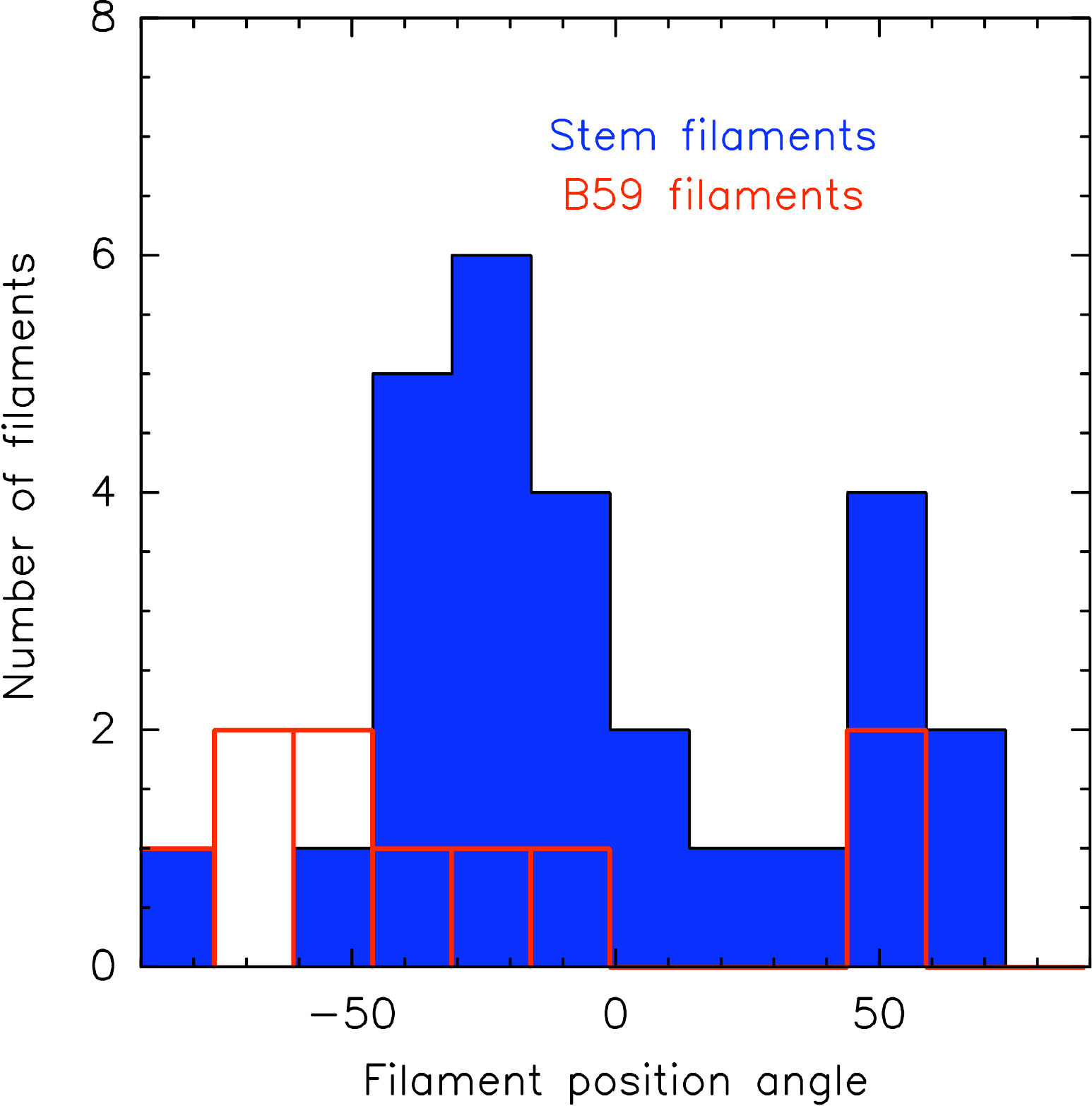}
       \caption{Histograms of position angles for the selected filaments: Stem filaments in blue, B59 filaments in red. A Kolmogorov-Smirnov test indicates there is only a $\sim10\%$ chance that these two sets of filaments have the same parent distribution of position angles.}
         \label{histo_pa}
   \end{figure}

The identified network of filaments shows some interesting features (see online Fig.~\ref{zoom}). While the Stem filaments are organized in a regular, grid-like, pattern, the B59 network is organized differently, with filaments converging toward the centre of the clump, in a manner reminiscent of a hub-like clump \citep{myers2009}. This can be better visualized by measuring the position angle (P.A.) of each of the red and cyan filaments in Fig.~\ref{skel}, and constructing the histograms of filament P.A. for both B59 and Stem regions. Figure~\ref{histo_pa} shows that there are mostly two preferential orientations in the Stem, at ${\rm P.A.} \simeq-25\degr$ and ${\rm P.A.}\simeq +55\degr$, nearly 90\degr\ from each other. No such trend is observed in B59, despite the low number statistics. A Kolmogorov-Smirnov test shows that there is only a $\sim10\%$ chance that the two samples of filaments in the Stem and in B59 have the same parent distribution of position angles.

We derived average radial column density profiles  for all identified filaments, as shown in Fig.~\ref{filprof} for two examples.
As in \citet{arzoumanian2011}, we fitted the inner part of the background-subtracted filament profiles with a Gaussian function. Based on these {Gaussian fits,} we estimate an average deconvolved FWHM width for the Pipe filaments of $0.08 \pm0.02$~pc (see online Fig.~\ref{histowidth}), close to the characteristic value of 0.1~pc found by \citet{arzoumanian2011} in other nearby interstellar clouds observed as part of the \emph{Herschel} GBS. In the Pipe, the majority of filaments are subcritical, with $M_{line} < M_{line,crit}$. Most importantly, we also find several filaments showing clear asymmetric transverse profiles. Asymmetric profiles are found at different locations along the Pipe complex, i.e. near the bow-like edge of B59, but also in the Stem. For low-density filaments, the background definition is important. In this paper, we chose a constant background at the level of the weaker profile wing (such as in Fig.~\ref{filprof}) since it is the one illustrating the best the asymmetry in the profiles. In practice, this background definition has little influence (less than a factor of 2) on the filament widths we measure\footnote{Three different background baseline can be considered: a constant background at the level of the weaker profile wing (which we adopt in this paper), a constant background at the level of the higher  profile wing, and a linearly interpolated background making the connection between the two profile wings. We performed three different fits for these three different backgrounds. For instance, for filament 1 the three fits provide FWHM widths ranging from 0.06~pc to 0.13~pc, while for filament 2 the derived widths ranges from 0.049~pc to 0.054~pc.}. To quantify this asymmetry, we calculated an asymmetry parameter, defined as the ratio of the integrated column density on either side of the profile peak (e.g., the blue shaded area over the yellow shaded area in Fig.~\ref{filprof}). We integrated over the same range of radii on either side of the filament crests (maximum of 0.2~pc, i.e. the default value). However, this range of radii can differ from filament to filament according to the scatter in the wings of the profiles (due, e.g., to nearby cores or filaments).  Figure~\ref{filorien} shows this column density ratio as a function of the position angle of the filaments. We clearly see that the most asymmetric filaments are oriented around ${\rm P.A.} =-5\degr$  (i.e., nearly vertical in Fig.~\ref{skel}), with higher column density found on their eastern sides (e.g. , for filaments with P.A $\sim 0\degr$, the blue shaded area in Fig.~\ref{filprof} lies east - cf Appendix).  This trend is only observed around this specific filament orientation.

\section{Discussion}

\subsection{Filament formation by large-scale compression ?}

The asymmetric filament profiles discussed above (cf. Fig.~\ref{filprof}), along with the presence of a bow-like edge of B59, suggest that several of the filamentary structures have been at least influenced, and possibly formed, by a compression flow coming from the western side of the Pipe Nebula. Interestingly, \citet{degeus1992} and \citet{onishi1999} proposed that winds from the Sco OB2 association ($d=145~{\rm pc} \pm2~{\rm pc}$ \citealt{dezeeuw1999}), located 30~pc away from the Pipe, could be responsible for triggering star formation in B59.  \citet{onishi1999} also showed that the energetics from this association were sufficient to influence the Pipe Nebula density structure. The Sco OB2 association lies to the west of B59, away  from the Pipe main body  (see online Fig.~\ref{scoob2}).  It is reasonable to propose that the asymmetric filaments are (being) formed by a compression powered by the Sco OB2 association.  
The interaction between this compressive flow and the Pipe provides a unique opportunity to study the complex mechanism of filament formation. 

Turbulence-generated compressive flows have been proposed to be the main driver of filament formation in the interstellar medium \citep{padoan2001,arzoumanian2011}. The chaotic nature of turbulence, however,  makes detailed analysis of such compressed filaments difficult. In the case of the Pipe,  we see this compression as a special configuration of turbulent shocks. Here, a shock front is travelling along a well-defined direction, as opposed to many shock fronts with random directions, simplifying  the problem.  The fact that the widths of the Pipe filaments (see Sect.~4 and online Fig.~\ref{histowidth}) match the characteristic value of 0.1~pc found by \citet{arzoumanian2011} in other nearby interstellar clouds supports the idea that wind compression and turbulent compression are closely related mechanisms. Note, however, that the Sco OB2 winds can compress material only on the surface of the cloud. As opposed to turbulent compression, these winds cannot therefore be responsible for the formation of filaments lying inside the cloud volume.

\begin{figure}
   \vspace{-0cm}
   \hspace{-0cm}
   \includegraphics[width=9cm,angle=0]{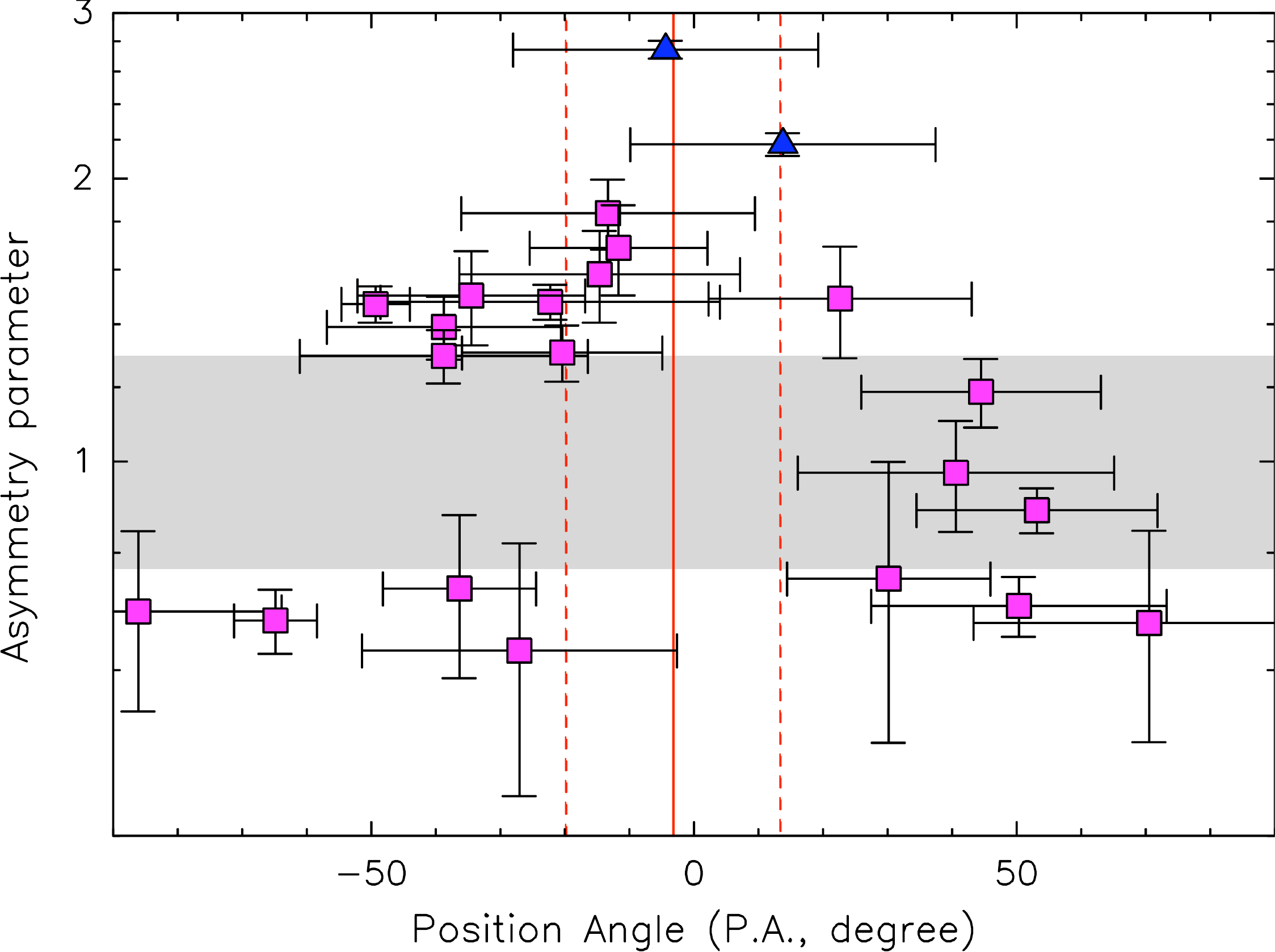}
   \vspace{-0.3cm}
      \caption{Filament position angle versus degree of asymmetry (see text and Fig.~\ref{filprof}). Only filaments with asymmetry parameter errors  $<30\%$ and P.A. uncertainties $<30\degr$ are shown.  The blue triangles correspond to the two filaments shown in Fig.~\ref{filprof}. Filaments with $-40\degr<{\rm P.A.}< +10$\degr\  are more asymmetric with more material towards the eastern part of the Pipe (i.e. away from the direction of the Sco OB2 association).  The shaded area shows the location of symmetric profiles (within a 30\% variation from symmetry). The vertical solid red line corresponds to the average orientation of the magnetic field in the region and the vertical dashed lines mark a $\pm 1\sigma$ deviation of the magnetic field angles.}
         \label{filorien}
   \end{figure}

As illustrated in Figs.~\ref{filorien} and \ref{cdmap}, the direction of the magnetic field \citep{franco2010} is nearly parallel to the compressed filaments observed in the Stem. While such similar orientations could be coincidental, in the following, we briefly discuss a mechanism which may be partly responsible for aligning the magnetic field lines  and the compressed filaments. Under the assumption of magnetic flux conservation, compression amplifies the magnetic field component parallel to a compressed layer but leaves the component perpendicular to the layer unaffected. As a consequence, the magnetic field lines will become nearly  parallel to the layer in the case of significant cloud compression. Assuming a plane-parallel compression, the combination of mass and magnetic flux conservation implies that the magnetic field amplification in compressed layers or filaments  can be expressed as 
$A_B = B^{//}_{post}/B^{//}_{pre} = n_{post}/n_{pre}$, 
where the {\it post} and {\it pre} suffixes stand for post-shock and pre-shock quantities, {\it B$^{//}$} is the strength of the magnetic field component parallel to the shock front, and {\it n} is the volume density. The pre-shock volume density can be approximated by the average volume density in the Pipe complex, i.e., $\sim 10^3$~cm$^{-3}$~\citep{onishi1999}. The post-shock volume density is given by the density in the central part of the filaments which, based on the present {\it Herschel} data, is estimated to be $\sim 1\times10^4$~cm$^{-3}$. These values suggest  $A_B\simeq 10$. Such an amplification is  sufficiently large that it may be partly responsible for the observed alignment between magnetic field lines and the compressed filaments.

We might expect that compression of a molecular cloud by a large-scale flow will form shells rather than filaments. The fact that the observed width of filaments is uniform (cf. online Fig.~\ref{histowidth}), however, favours the filament geometry. In addition, theoretical arguments suggest that shells quickly fragment into filamentary structures. As argued by \citet{nagai1998}, when the ram pressure of a compressive flow dominates the pressure within a magnetized shocked layer, then the layer will fragment into a set of filaments whose main axes lie within the layer, parallel to the amplified component of the magnetic field. Conversely, when the compressed layer is self-gravitating, it will fragment in a direction perpendicular to the magnetic field. This mechanism would lead to the formation of bound or unbound filaments whose masses per unit length are above or below, respectively, the critical value of $\sim15$~M$_{\odot}$\,pc$^{-1}$ (or, equivalently, a column density of $\sim7\times10^{21}$~cm$^{-2}$ for 0.1~pc-wide filaments, see Sect.~1). This mechanism could also explain why some of the Pipe filaments, whose column densities are below this critical value (online Fig.~\ref{cdmap}),  are roughly aligned with the magnetic field.


\subsection{Filaments and star formation in B59}


B59 differs from the rest of the Pipe Nebula. It is the only portion of the complex which is forming stars
 \citep{forbrich2009}. 
 In addition, the bow-like front and the sharp edge on the western side of B59 provide a strong indication that it is interacting with a large-scale external compressive flow coming from the west. We propose that such a compression may have enhanced, in a "snow plough" fashion, the column density contrast of the original B59 clump, rendering it prone to global gravitational collapse \citep[cf.][]{peretto2006}. This scenario is consistent with the different timescales involved. The Sco OB2 association has an estimated age of $\sim 5$~Myr \citep{degeus1989} and the B59 cluster an age of $\le2.6$~Myr \citep{covey2010}. These ages are consistent with a sequential star formation scenario in which the Sco OB2 wind compression has triggered star formation in B59.  Note that the star formation activity in Ophiuchus (online Fig.~\ref{scoob2}) has also likely been triggered by the  Sco OB2 winds \citep{nutter2006}.

Finally, the different patterns between the nearly grid-like filament network of the Stem and the centrally converging network of B59 (see Sect.~3 and online Fig.~\ref{zoom}) may be interpreted as a consequence of two different physical regimes. The Stem filamentary structures appear to be mostly subcritical (with $M_{line} < M_{line, crit}$) and show a pattern reflecting the initial conditions of their formation. In contrast, the gravity dominates in B59 (Franco et al. 2010, see also online Fig.~\ref{cdmap}) and may have therefore organized the filamentary structures in a pattern converging toward the center of mass of the contracting protocluster. Such hub-like structures are observed in many other star-forming regions (Myers 2009, see also Schneider et al. 2012 for another \emph{Herschel} example), and might point toward a common pattern of gravitationally unstable clumps. Local feedback from collimated protostellar outflows, however, may also have influenced, on smaller scales, the morphology of some of the observed B59 filaments \citep{cabral2012}. While the question of the relative impact of large-scale compression, gravity and protostellar  outflows remains open in B59, we believe that large-scale compression is a plausible mechanism for the {\it initial} formation of filamentary structures in the rest of complex.

\phantom{\citet{schneider2012}}

\begin{acknowledgements}
We thank the anonymous referee for his thorough report which helped improving the clarity
of this paper. N.P. is supported by a CEA/Marie Curie Eurotalents fellowship.
SPIRE has been developed by a consortium of institutes led by Cardff Univ. (UK) and 
including Univ. Lethbridge (Canada); NAOC (China); CEA, LAM (France); 
IFSI, Univ. Padua (Italy); IAC (Spain); Stockholm Observatory (Sweden); 
Imperial College London, RAL, UCL-MSSL, UKATC, Univ. Sussex (UK); 
Caltech, JPL, NHSC, Univ. Colorado (USA). This development has been sup- 
ported by national funding agencies: CSA (Canada); NAOC (China); CEA, 
CNES, CNRS (France); ASI (Italy); MCINN (Spain); SNSB (Sweden); STFC 
and UKSA (UK); and NASA (USA). PACS has been developed by a consortium of in- 
stitutes led by MPE (Germany) and including UVIE (Austria); KUL, CSL, 
IMEC (Belgium); CEA, OAMP (France); MPIA (Germany); IFSI, OAP/AOT, 
OAA/CAISMI, LENS, SISSA (Italy); IAC (Spain). This development has been 
supported by the funding agencies BMVIT (Austria), ESA-PRODEX (Belgium), 
CEA/CNES (France), DLR (Germany), ASI (Italy), and CICT/MCT (Spain). 
\end{acknowledgements}

\bibliographystyle{aa}
\bibliography{references}

\begin{appendix}

\section{Angle definition and filament orientation measurement}

All angles in this paper are defined in the same frame,  using the standard definition of position angles, measured east of north, i.e. 0\degr\ toward the north and positive angles counterclockwise. The orientation of filaments is measured between -90\degr\ and +90\degr\ . Tangent angles are first calculated along the crest of each filament. The median value of these tangent angles provides the filament orientation while the absolute deviation around the filament orientation provides the angle uncertainty used in Fig.~\ref{filorien}. 

\begin{figure}
   \vspace{-0.0cm}
   \hspace{.5cm}
   \includegraphics[width=7.5cm,angle=0]{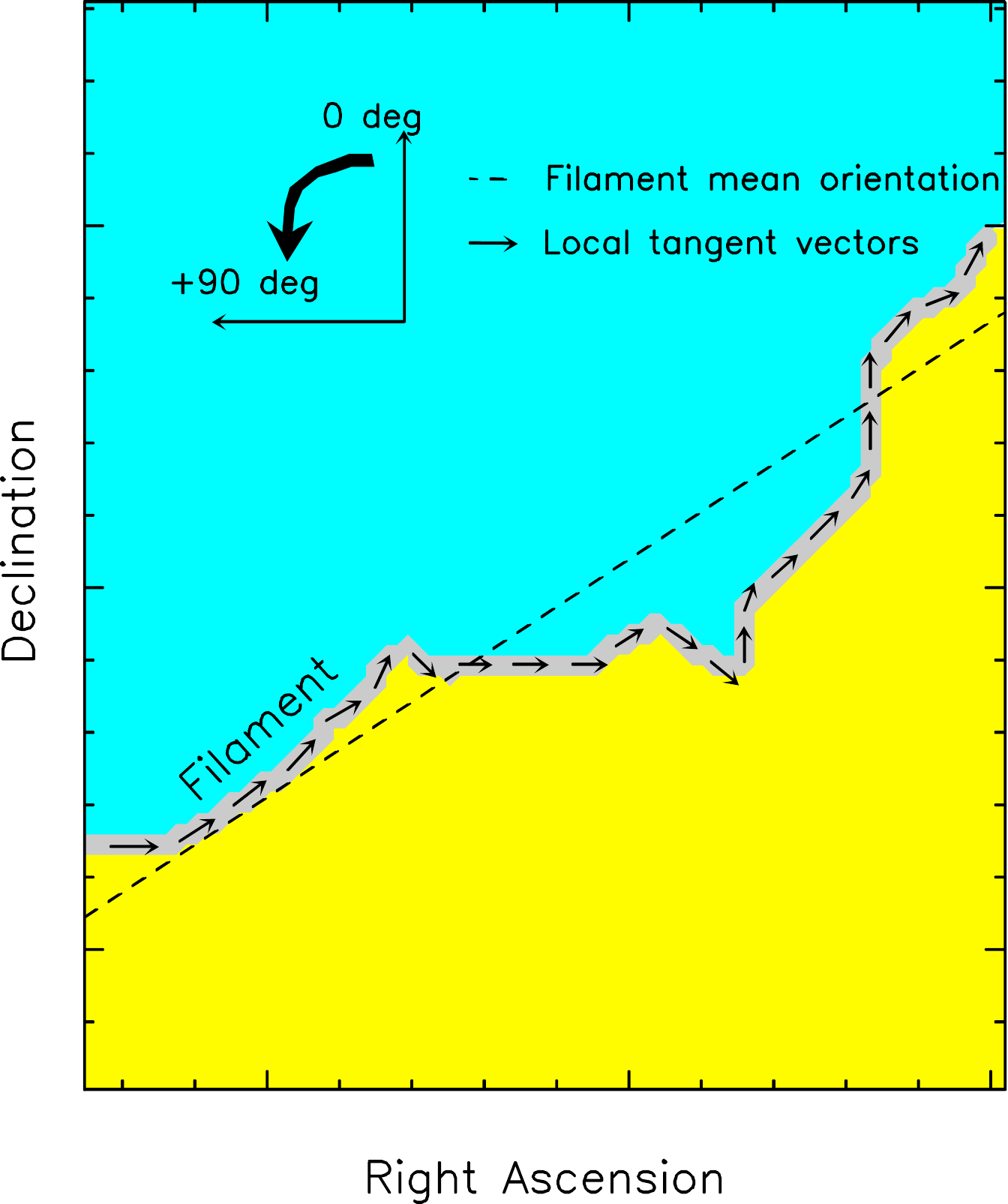} 
   \vspace{-0cm}
      \caption{Illustration of filament position angles, measured east of north, and tangent vectors along a particular filament. Local column density profiles are calculated for each pixel along the crest of the filament, in the direction perpendicular to the tangent vectors. If the local perpendicular direction is closer than 45\degr\ from the filament mean orientation, the corresponding local profile is not considered when averaging the column density profiles.
      }
         \label{filex}
   \end{figure}

To construct average column density profiles such as those shown in Fig.~\ref{filprof}, we first constructed a series of profiles, one per pixel along the filament crest, in the direction perpendicular to the local tangent. 
 Before averaging these local profiles together, we have to ensure that they are similarly oriented (e.g. east part of each local profile averaged together). For this purpose, we followed each filament in the direction of the filament orientation, always starting by constructing the local profile toward positive angles, i.e. counterclockwise. The asymmetry parameter was always calculated by integrating the positive side (blue shaded area on Fig.~\ref{filex}) over the negative one (yellow shaded area). This convention implies that for a filament position angle of 0\degr, an asymmetry parameter above 1 means that the eastern side exhibits higher column density.

\end{appendix}

\Online


\onlfig{5}{
  \begin{figure*}
   \vspace{-0cm}
   \hspace{0.5cm}
   \includegraphics[width=16.5cm,angle=0]{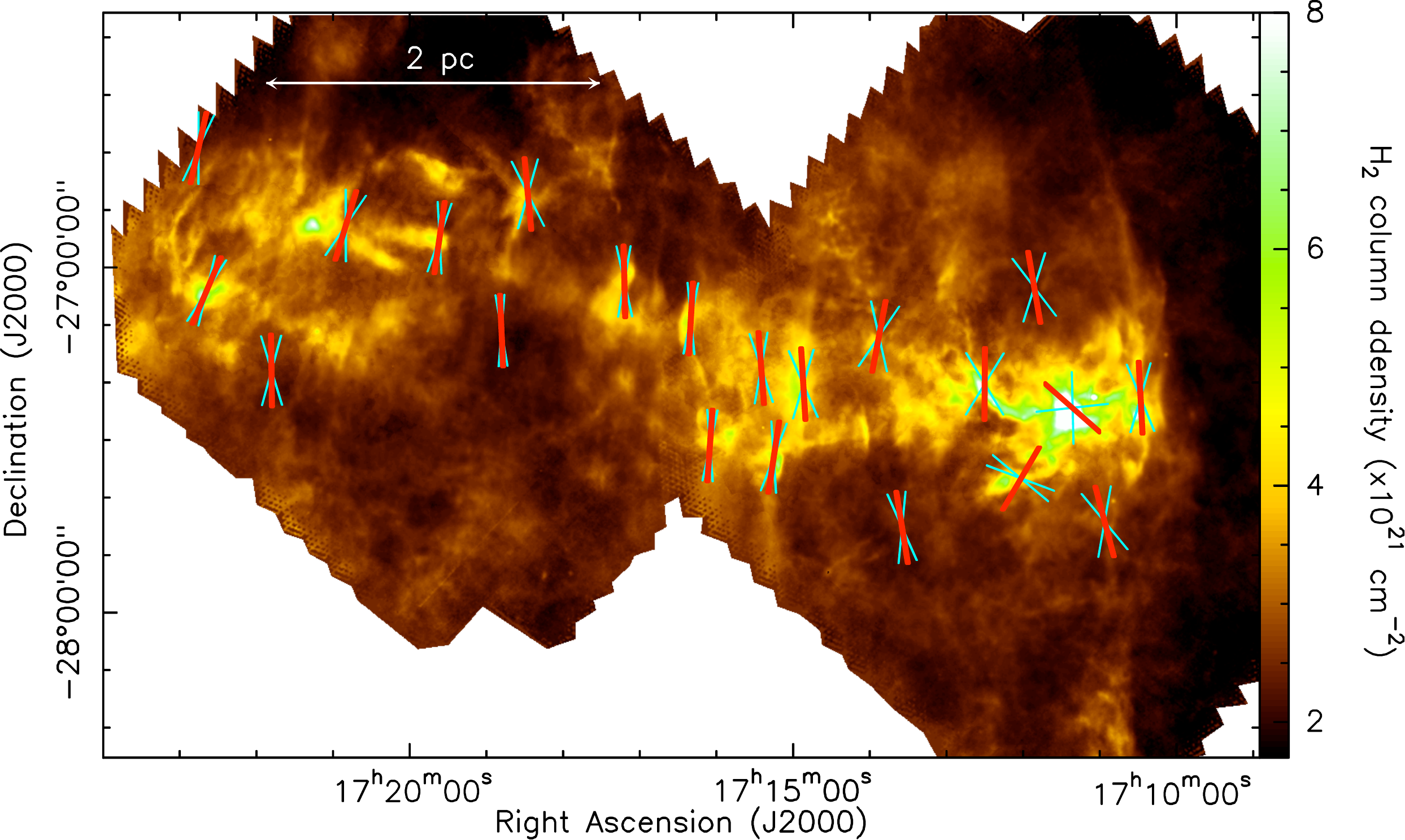} 
   \vspace{-0cm}
      \caption{\emph{Herschel} column density map of the B59 and Stem in the Pipe complex, as presented in Fig.~\ref{skel}. 
 The red segments provide the direction of the local magnetic field as measured from near-infrared dust polarisation data \citep{franco2010}. The blue segments mark a  $\pm2\sigma$ deviation on either side of the nominal field direction. Note that gravitationally unstable regions, those lying above $N_{H_2}\simeq 7\times10^{21}$~cm$^{-2}$, correspond to the white areas. 
              }
         \label{cdmap}
   \end{figure*}
}

\onlfig{6}{
 \begin{figure*}
   \vspace{-0.0cm}
   \hspace{-0.1cm}
   \includegraphics[width=8.5cm,angle=0]{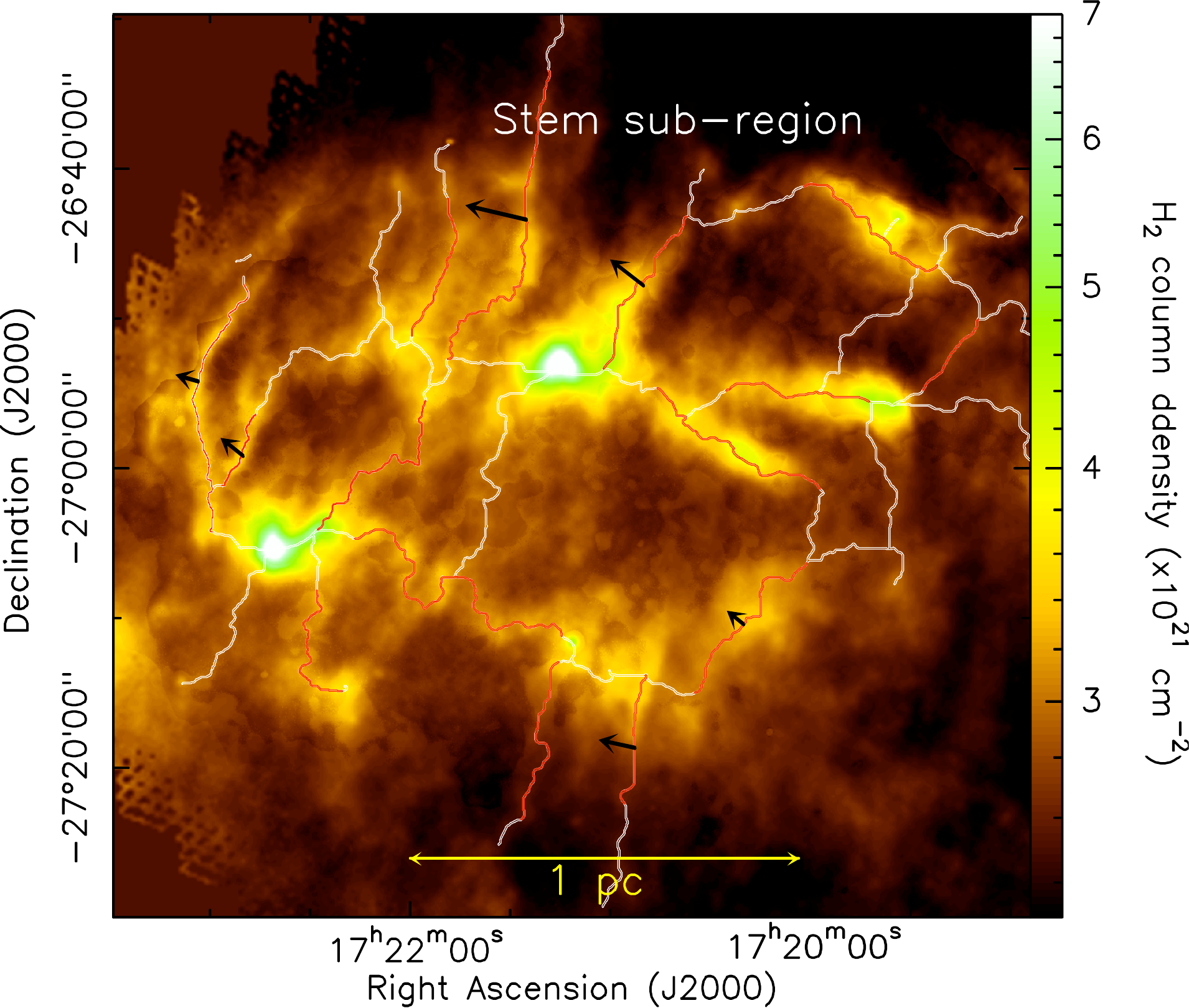} 
   \hspace{0.5cm}
   \includegraphics[width=8.5cm,angle=0]{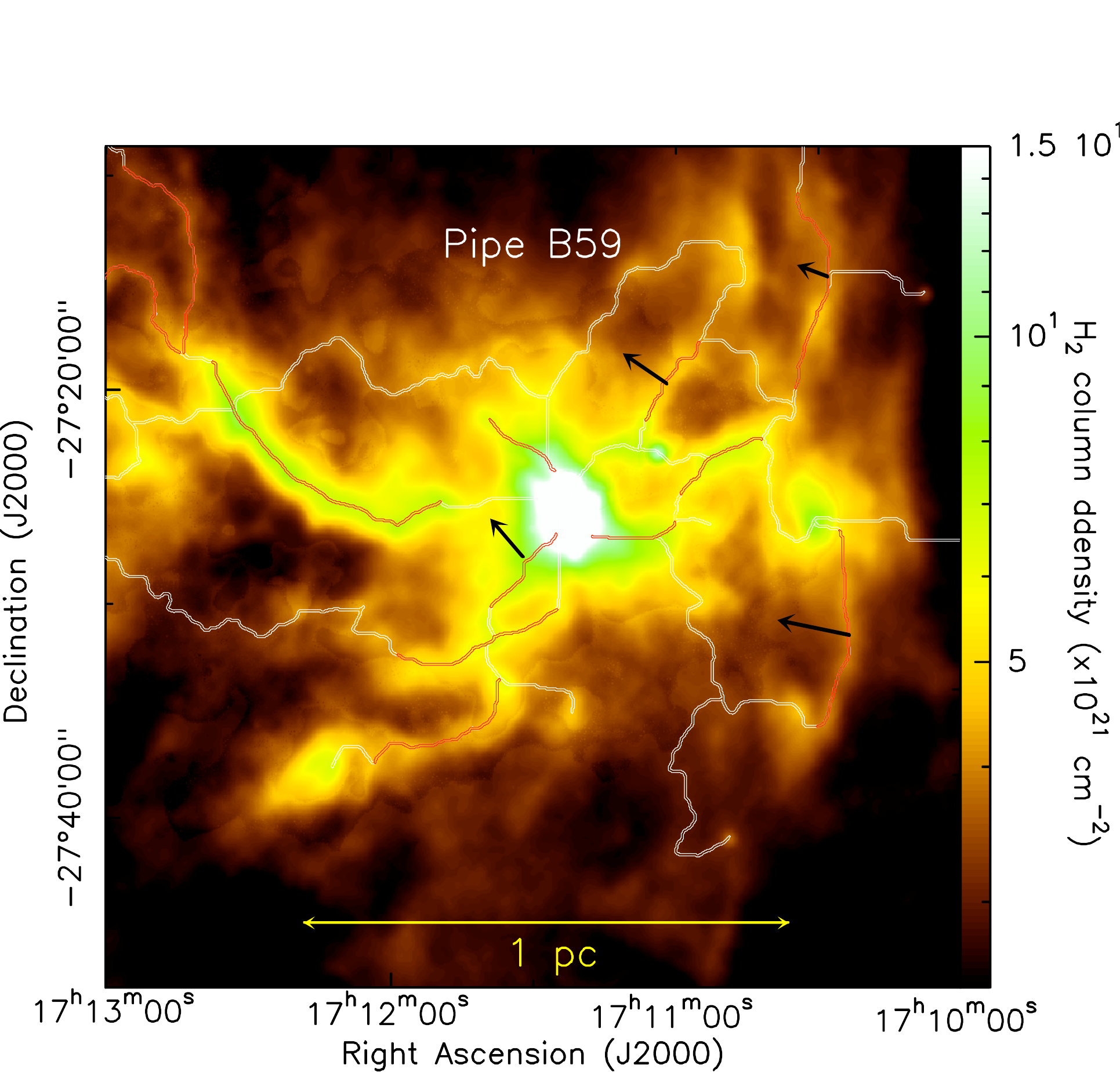} 
   \vspace{-0cm}
      \caption{ Blow-ups of Fig.~\ref{skel} showing the column density images of a portion of the Stem (left) and the B59 (right) regions over which the filament networks as traced by the DisPerSE algorithm are plotted. On these images, we can see better the  "grid-like"  pattern in the Stem, and the converging pattern of B59. The direction and length of each arrow show the direction and amplitude, respectively,  of the asymmetry in the corresponding column density profile, for filaments having $-60\degr<P.A.<+30\degr$ and asymmetry uncertainty better than 10\%.
      }
         \label{zoom}
   \end{figure*}
}

\onlfig{7}{
\begin{figure*}
   \vspace{-0cm}
   \hspace{5cm}
   \includegraphics[width=7.8cm,angle=0]{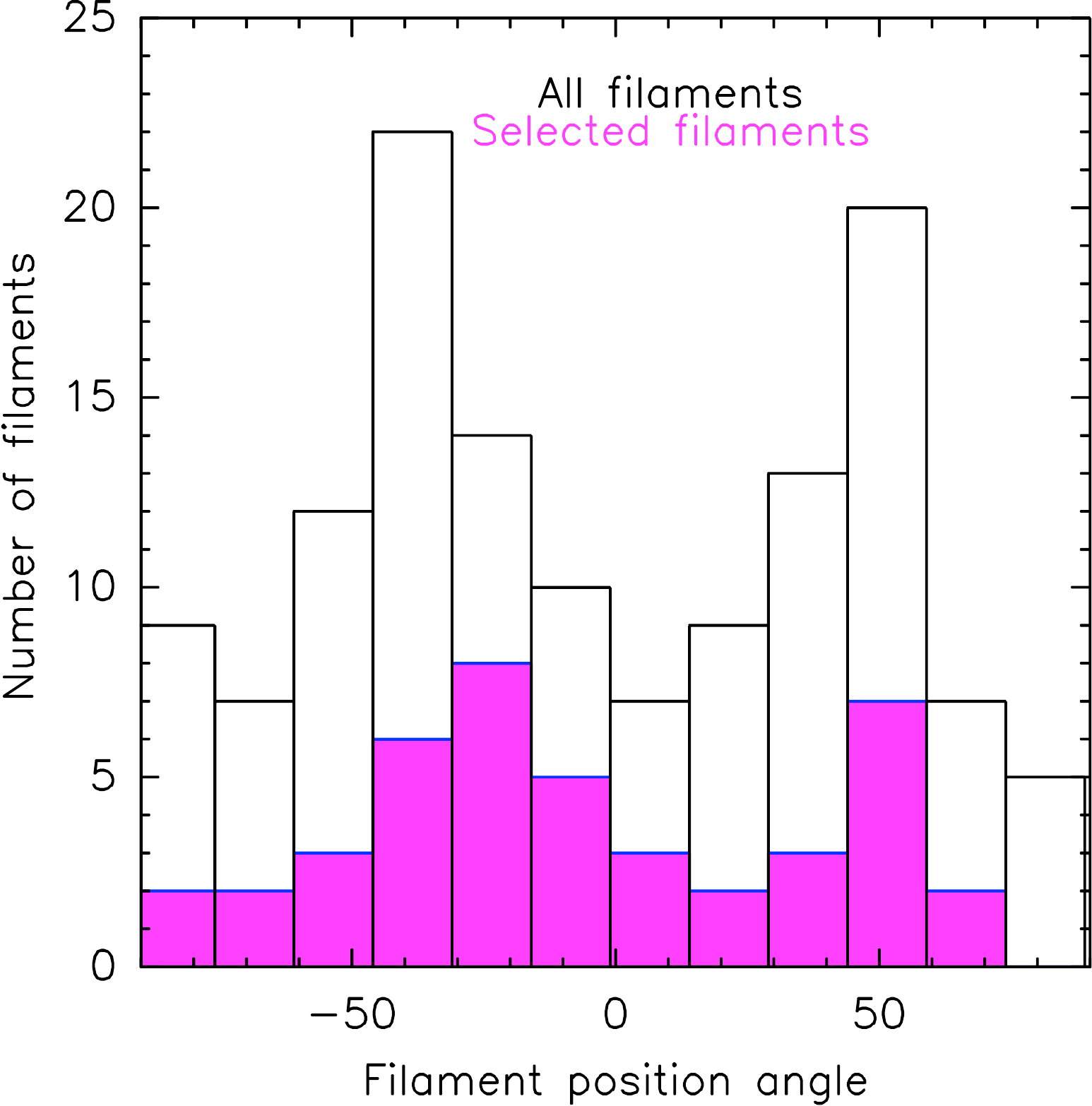}
       \caption{Histograms of position angles for all the filaments identified with DisPerSE (solid black line) and the selected filaments (purple). A Kolmogorov-Smirnov test indicates a $\sim87\%$ chance that these two sets of filaments have the same parent distribution of position angles.}
         \label{histo_pa_all}
   \end{figure*}
}

\onlfig{8}{
 \begin{figure*}
   \vspace{-0.5cm}
   \hspace{4cm}
   \includegraphics[width=10cm,angle=0]{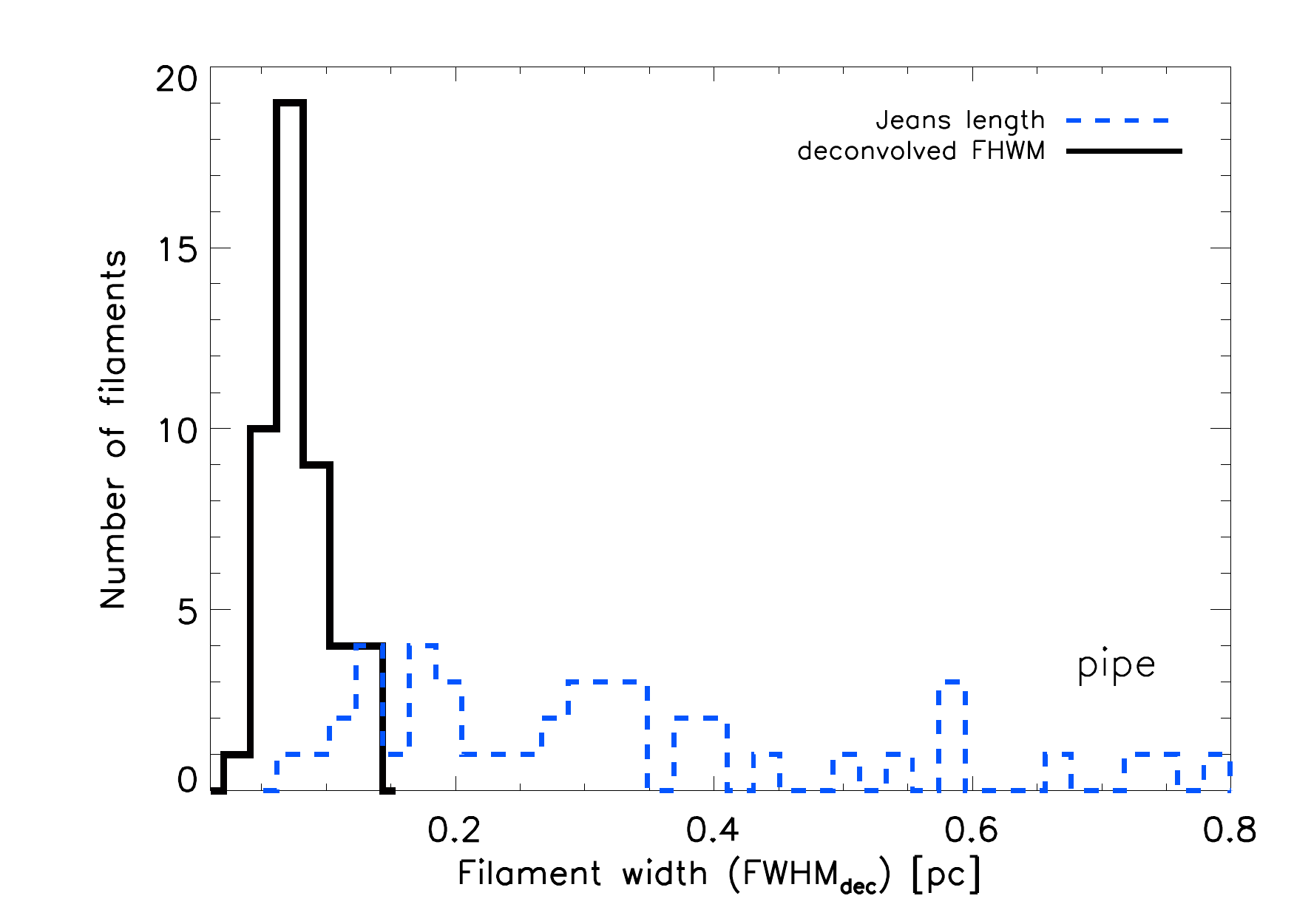} 
   \vspace{-0cm}
      \caption{Histogram of the deconvolved FWHM widths (black solid line) of the filaments identified in the {\it Herschel} Pipe images presented in this paper. The histogram peaks at $\sim0.08$~pc and has a standard deviation of 0.02~pc. The blue dashed line represents the histogram of Jeans lengths measured from  the central column densities of the filament (see Arzoumanian et al. 2011).
      }
         \label{histowidth}
   \end{figure*}
}

\onlfig{9}{
 \begin{figure*}
   \vspace{-0cm}
   \hspace{5cm}
   \includegraphics[width=8cm,angle=0]{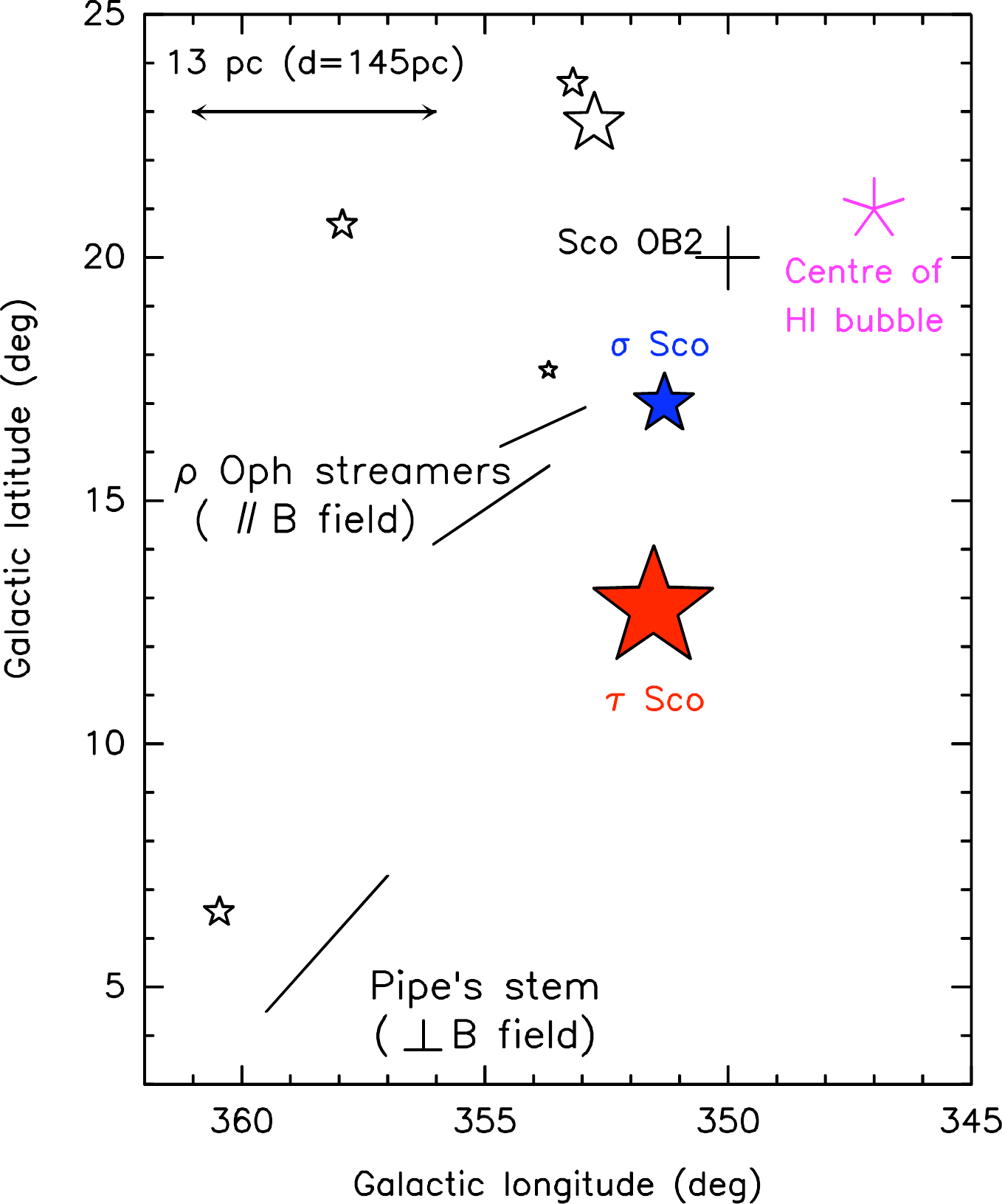}
   \vspace{-0cm}
      \caption{Schematic view of the Pipe nebula cloud, Ophiuchus streamers and Sco OB2 association. The solid straight lines represent the streamers and main body of the Ophiuchus and Pipe molecular clouds, respectively. The plus symbol marks the central position of the Sco OB 2 association. The star symbols represent the positions of some of the main members of the association, and the HI bubble observed by \citet{degeus1992} is also indicated.
      }
         \label{scoob2}
   \end{figure*}
}

\end{document}